\documentclass[aps,pra,twocolumn,notitlepage,superscriptaddress]{revtex4-1}

\usepackage{physics}
\usepackage{xcolor,graphicx}
\usepackage{comment}
\usepackage{siunitx}

\usepackage{amsthm,amssymb}
\usepackage{array}
\usepackage{hyperref}
\usepackage{mathtools}
\usepackage{bm,times,enumitem}
\usepackage{physics}
\usepackage{algorithm}
\usepackage{algc}
\usepackage{algcompatible}
\usepackage{algpseudocode}
\usepackage{soul}

\begin{document}

\title{Non-asymptotic Heisenberg scaling: experimental metrology for a wide resources range}

\author{Valeria Cimini}
\affiliation{Dipartimento di Fisica, Sapienza Universit\`{a} di Roma, Piazzale Aldo Moro 5, I-00185 Roma, Italy}

\author{Emanuele Polino}
\affiliation{Dipartimento di Fisica, Sapienza Universit\`{a} di Roma, Piazzale Aldo Moro 5, I-00185 Roma, Italy}

\author{Federico Belliardo}
\affiliation{NEST, Scuola Normale Superiore and Istituto Nanoscienze-CNR, I-56126 Pisa, Italy}

\author{Francesco Hoch}
\affiliation{Dipartimento di Fisica, Sapienza Universit\`{a} di Roma, Piazzale Aldo Moro 5, I-00185 Roma, Italy}

\author{Bruno Piccirillo}
\affiliation{Department of Physics ``E. Pancini'', Universit\'a di Napoli "Federico II", Complesso Universitario MSA, via Cintia, 80126, Napoli}

\author{Nicol\`o Spagnolo}
\affiliation{Dipartimento di Fisica, Sapienza Universit\`{a} di Roma, Piazzale Aldo Moro 5, I-00185 Roma, Italy}

\author{Vittorio Giovannetti}
\email{vittorio.giovannetti@sns.it}
\affiliation{NEST, Scuola Normale Superiore and Istituto Nanoscienze-CNR, I-56126 Pisa, Italy}

\author{Fabio Sciarrino}
\email{fabio.sciarrino@uniroma1.it}
\affiliation{Dipartimento di Fisica, Sapienza Universit\`{a} di Roma, Piazzale Aldo Moro 5, I-00185 Roma, Italy}

\begin{abstract}

Adopting quantum resources for parameter estimation discloses the possibility to realize quantum sensors operating at a sensitivity beyond the standard quantum limit. Such approach promises to reach the fundamental Heisenberg scaling as a function of the employed resources $N$ in the estimation process. Although previous experiments demonstrated precision scaling approaching Heisenberg-limited performances, reaching such regime for a wide range of $N$ remains hard to accomplish. Here, we show a method which suitably allocates the available resources reaching Heisenberg scaling without any prior information on the parameter. We demonstrate experimentally such an advantage in measuring a rotation angle. We quantitatively verify Heisenberg scaling for a considerable range of $N$ by using single-photon states with high-order orbital angular momentum, achieving an error reduction greater than $10$ dB below the standard quantum limit. Such results can be applied to different scenarios, opening the way to the optimization of resources in quantum sensing.

\end{abstract}

\maketitle

The measurement process permits to gain information about a physical parameter at the expense of a dedicated amount of resources $N$. Intuitively, the amount of information that can be extracted will depend on the number of employed resources, thus affecting the measurement precision on the parameter. By limiting the process to using only classical resources, the best achievable sensitivity is bounded by the standard quantum limit (SQL) and it scales as $1/\sqrt{N}$. Such limit can be surpassed by employing $N$ quantum resources, defining the ultimate precision bound $\pi/N$, known as the Heisenberg limit (HL) \cite{berry2001optimal,PhysRevLett.124.030501}. To achieve such fundamental limit \cite{Giovannetti1330, Giovannetti,giovannetti2006quantum}, a crucial requirement is the capability of allocating them efficiently. Indeed, the independent use of each resource results in an uncertainty which scales as the SQL, while the optimal sensitivity can be achieved exploiting quantum correlations in the probe preparation stage \cite{lee2002quantum,bollinger1996optimal}.

An example of quantum resource enabling Heisenberg-limit performances in parameter estimation is the class of two-mode maximally entangled states, also called N00N states. Such kind of states have been widely exploited in quantum metrology experiments performed on photonic platforms \cite{avsreview2020}. In particular, one of the most investigated scenarios is the study of the phase sensitivity resulting from interferometric measurements, thanks to their broad range of applications ranging from imaging \cite{PhysRevLett.85.2733} to biological sensing \cite{Wolfgramm2013,Cimini:19}. In this context, the optimal sensitivity can be achieved through the super-resolving interference obtained with $N$ photons N00N states \cite{Mitchell,avsreview2020}. However, current experiments relying on N00N states are limited to regimes with small number of $N$ \cite{Nagata726,Daryanoosh2018,Roccia:18,PhysRevLett.112.223602,Afek879}. Indeed, scaling the number of entangled particles in such kind of states is particularly demanding due to the high complexity required for their generation, that cannot be realized deterministically with linear optics for $N>2$. Experiments with up to ten-photon states have been realized \cite{PhysRevLett.117.210502,PhysRevA.85.022115}, but going beyond such order of magnitude requires a significant technological leap. Furthermore, this class of states results to be very sensitive to losses, which quickly cancels the quantum advantage as a function of the number of resources $N$. For this reason, the unconditional demonstration of a sub-SQL estimation precision, taking into account all the effective resources, has been reported only recently in Ref. \cite{Slussarenko2017} with two-photon states. 

Alternative approaches have been implemented for \emph{ab-initio} phase estimations, sampling multiple times the investigated phase shift \cite{Higgins_2009} through adaptive and non-adaptive multi-pass strategies \cite{PhysRevA.63.053804,Berni2015,Higgins_2009}, achieving the HL in an entanglement-free fashion. However, one of the main challenges is to maintain the Heisenberg scaling when increasing the number of dedicated resources. Beyond the experimental difficulties encountered when increasing the number of times the probe state propagates through the sample, such protocols become exponentially sensitive to losses. Therefore, the demonstration of Heisenberg-limited precision with such an approach still remains confined to small $N$.

All previous approaches present a fundamental sensitivity to losses, which prevents the observation of Heisenberg limited performances in the asymptotic limit of very large $N$ where the advantage substantially reduces to a constant factor \cite{Escher2012}. Thus, it becomes crucial to focus the investigation of quantum-enhanced parameter estimation in the non-asymptotic regime, with the aim of progressively extending the range of observation of Heisenberg scaling sensitivity (in $N$). To this end, it is necessary to properly allocate the use of resources in the estimation process. In this Article, using N00N-like quantum states encoded in the total angular momentum of each single photon, more robust to losses than the aforementioned approaches, we implement a method able to identify and implement optimal allocation of the available resources. We test the developed protocol for an \emph{ab-initio} measurement of a rotation angle in the system overall periodicity interval $[0,\pi)$, resolving the ambiguity among the possible equivalent angle values. We perform a detailed study on the precision scaling as a function of the dedicated resources, demonstrating Heisenberg limited performances for a wide range of the overall amount of resources $N$.

\section*{Protocol}

\begin{figure*}[t]
  \includegraphics[width=\textwidth]{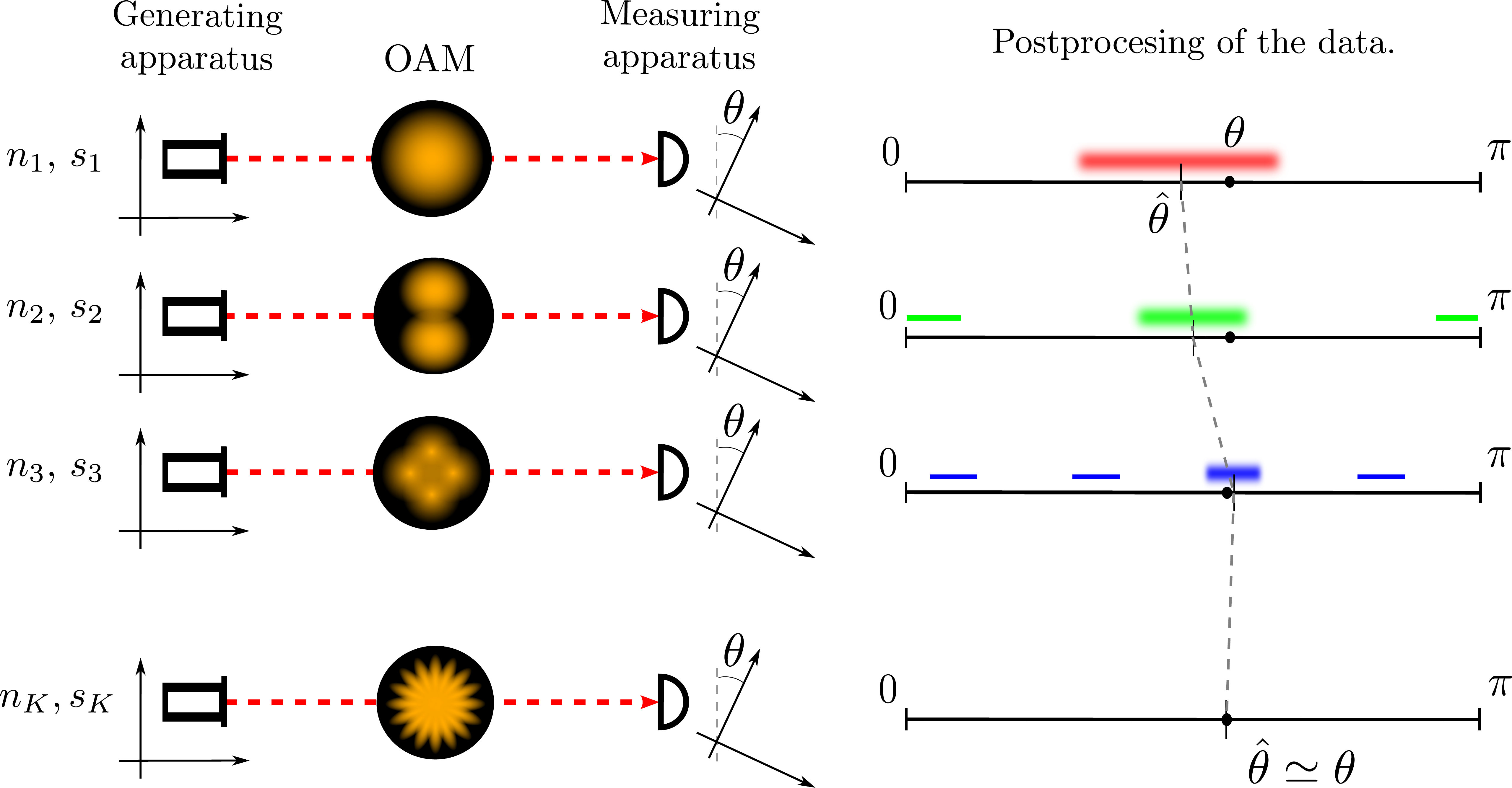}
    \caption{{\bf Conceptual scheme of the estimation protocol.} At stage $i$, {$n_i$ copies of the state $\ket{\Psi_{s_i}(\theta)}$ characterized by a resource number $s_i$
are employed.} In our experiment the encoded parameter is the rotation angle $\theta$ between two reference systems, associated respectively to two platforms corresponding to the photon generation stage and to the measurement apparatus. The quantum resource is related to the total angular momentum of the photons. Each stage is successful if and only if it could identify the correct interval for the angle $\theta$ among the possible ones, using the information on the previously selected interval. This allows the last stage (with maximal sensitivity) to produce an unambiguous estimator $\hat{\theta}$. In the figure the plausible intervals for the phase, computed from the outcomes of the independent and non-adaptive measurements are colored, and the selected one is highlighted.}
\label{fig:schema}
\end{figure*}

In a typical optical quantum estimation scheme one is interested in recovering the value of an unknown parameter $\theta\in [0, \pi)$ represented by an optical phase shift or, as in the case discussed in this work, by a rotation angle between two different platforms. The idea is then to prepare a certain number of copies $n$ of the input state $(\ket{0} + \ket{1})/\sqrt{2}$, let transform each one of them into the associated output configuration $\ket{\Psi_s(\theta)} = (\ket{0} + e^{-i2 s \theta} \ket{1})/\sqrt{2}$ by a proper imprinting process, and then measure, see Fig.~\ref{fig:schema}. In these expressions $\ket{0}$ and $\ket{1}$ stand for proper orthogonal states of the e.m. field. The integer quantity $s$ describes instead the amount of quantum resources devoted in the production of each individual copy of $\ket{\Psi_s(\theta)}$, i.e., adopting the language of Ref.~\cite{giovannetti2006quantum}, the number of {\it black-box operations} needed to imprint $\theta$ on a single copy of $(\ket{0} + \ket{1})/\sqrt{2}$. Therefore, in the case of $n$ copies, the total number of operations corresponds to  $ns$. For instance, in the scenario where one has access to a joint collection of $s$ correlated modes which get independently imprinted by $\theta$, $\ket{0}$ can be identified with the joint vacuum state of the radiation and $\ket{1}$ with a tensor product Fock state where all the modes of the model contain exactly one excitation (in this case $s$ can also be seen as the size  of the GHZ state $(\ket{0} + \ket{1})/\sqrt{2}$). On the contrary, in a multi-round scenario where a single mode undergoes $s$ subsequent imprintings of $\theta$, $\ket{0}$ and $\ket{1}$ represent instead the zero and one photon states of such mode.

The problem of determining the optimal allocation of resources that ensures the best estimation of $\theta$ is that, while states $\ket{\Psi_s(\theta)}$ with larger $s$ have greater sensitivity to changes in $\theta$, an experiment that uses just such output signals will only be able to distinguish $\theta$ within a period of size $\pi/s$, being totally blind to the information on where exactly locate such interval into the full domain $\left[0, \pi \right)$. The problem can be solved by using a sequence of experiments with growing values $s$ of the allocated quantum resources. We devised therefore a multistage procedure that works with an arbitrary growing sequence of $K$ quantum resources $s_1; s_2; s_3; \dots; s_K$, aiming at passing down the information stage by stage in order to disambiguate $\theta$ as the quantum resource (i.e. the sensitivity) grows, see Fig.~\ref{fig:schema} for a conceptual scheme of the protocol.

At the $i$-th stage $n_i$ copies of 
$\ket{\Psi_{s_i}(\theta)}$ are measured, individually and non-adaptively, and a multi-valued  ambiguous estimator is constructed. Then $s_i$ plausible intervals for the phase are identified, centered around the many values of the ambiguous multi-valued estimator. Finally, one and only one value is deterministically chosen according to the position of the selected interval in the previous step, removing the estimator ambiguity. At each stage the algorithm might incur in an error, providing an incorrect range selection for $\theta$. When this happens the subsequent stages of estimation are also unreliable. The probability of an error occurring at the $i$-th stage decreases as increasing the number $n_i$ of probes used in such a stage. The precision of the final estimator, $\hat{\theta}$, resulting from the multistage procedure is optimized in the number of probes $n_1, n_2, \dots, n_K$. The overall number of consumed resources, $N = \sum_{i=1}^K n_i s_i$, is kept constant. We thus obtain the optimal number of probes $n_i$ to be used at each stage. The details of the algorithm and the optimization are reported in the Methods. Remarkably, it can be analytically proved that such protocol with $s_i=2^{i-1}$ works at the Heisenberg scaling~\cite{Higgins_2009, kimmel_robust_2015, belliardo_achieving_2020}, provided that the right probe distribution is chosen. Due to the limited amount of available quantum resources, when growing the total number $N$ of resources, the scaling of the error $\Delta \hat{\theta}$ eventually approaches the SQL. In the non-asymptotic region, however, a sub-SQL scaling is reasonably expected. An important feature of such protocol is that, being non-adaptive, the measurement stage decouples completely from the algorithmic processing of the measurement record. This means that the algorithm producing the estimator $\hat{\theta}$ can be considered a post processing of the measured data. Non-unitary visibility can be easily accounted for in the optimization of the resource distribution. We emphasize that this phase estimation algorithm has been adapted to work for an arbitrary sequence of quantum resources, in contrast with previous formulations~\cite{higgins2007entanglement, Higgins_2009, kimmel_robust_2015}.

\section*{Experimental setup}
The total angular momentum of light is given by the sum of the spin angular momentum, that is, the polarization with eigenbasis given by the two circular polarizations states of the photon, and its orbital angular momentum (OAM). The latter is associated to modes with spiral wavefronts or, more generally, to modes having non-cylindrically symmetric wavefronts \cite{padgett2004light,erhard2018twisted}. The OAM space is infinite-dimensional and states with arbitrarily high OAM values are in principle possible. This enables to exploit OAM states for multiple applications such as quantum simulation \cite{cardano2016statistical,cardano_zak_2017,Buluta2009}, quantum computation~\cite{bartlett2002quantum,ralph2007efficient,Lanyon2009,Michael2016} and quantum communication~\cite{Wang2015,Mirhosseini_2015,krenn2015twisted,Malik2016,Sit17,Cozzolino2019_fiber,wang2016advances,cozzolino2019air}. Recently, photons states with more than $10,000$ quanta of orbital angular momentum have been experimentally generated \cite{Fickler13642}. Importantly, states with high angular momentum values can be also exploited to improve the sensitivity of the rotation measurements \cite{barnett2006resolution,jha2011supersensitive,fickler2012quantum,dambrosio_gear2013,Fickler2021}, thanks to the obtained super-resolving interference. The single-photon superposition of opposite angular momenta, indeed, represents a state with N00N-like features when dealing with rotation angles. Furthermore, the use of OAM in this context is more robust against losses compared both to approaches relying on entangled states or multi-pass protocols.

\begin{figure*}[ht!]
  \includegraphics[width=0.99\textwidth]{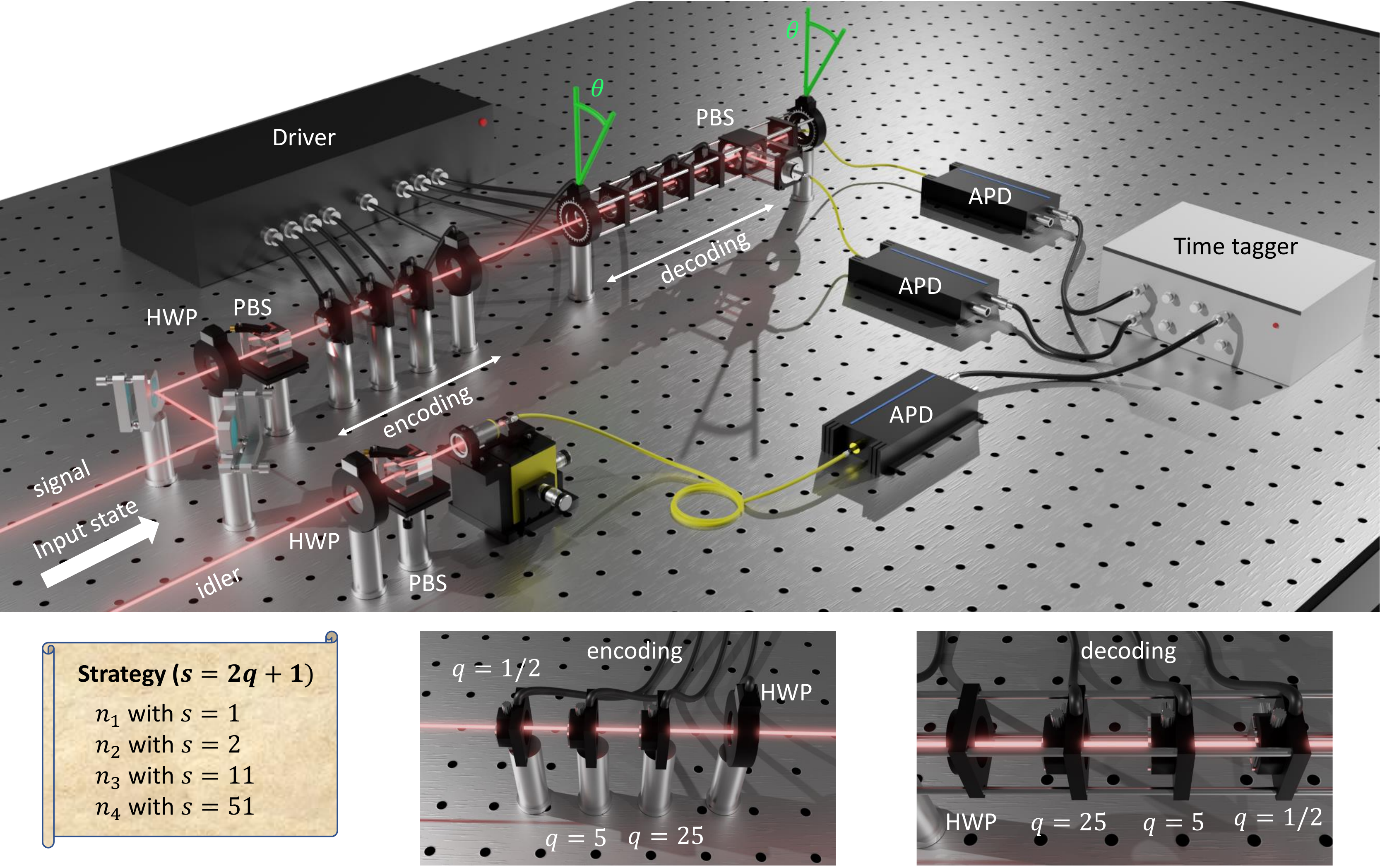}
    \caption{{\bf Experimental setup.} Single photons pairs are generated by a degenerate type-II SPDC process inside a ppKTP pumped by a $405$ nm cw laser. The idler photon is measured by a single photon detector (APD) and acts as a trigger for the signal that enters in the apparatus. This consists of an encoding stage which is composed of a first polarizing beam splitter (PBS) and three q-plates with different topological charges $q=1/2, 5, 25$, respectively, followed by a motorized half-waveplate (HWP). The decoding stage is composed by the same elements of the preparation mounted, in the reverse order, in a compact and motorized cage which can be freely rotated around the light propagation axis of an angle $\theta$. After the final PBS, the photons are measured through single photon detectors (APDs). Coincidences with the trigger photon are measured, analyzed via a time-tagger, and sent to a computing unit. The latter, according to the pre-calculated optimal strategy, controls all the voltages applied to the q-plates and the angle of rotation of the measurement stage.}
    \label{fig:setup}
\end{figure*}

In the present experiment, we employ the total angular momentum of single-photons as a tool to measure the rotation angle $\theta$ between two reference frames associated to two physical platforms \cite{dambrosio_gear2013}. The full apparatus is shown in Fig.~\ref{fig:setup}. The key elements for the generation and measurement of OAM states are provided by q-plates (QPs) devices, able to modify the photons OAM conditionally to the value of their polarization. A q-plate is a topologically charged half-wave plate that imparts an OAM $ 2\hbar\,q$ to an impinging photon and flips its handedness \cite{marrucci-2006spin-to-orbital}. 

In the preparation stage, single photon pairs at $808$nm are generated by a $20$mm-long periodically poled titanyl phosphate  (ppKTP) crystal pumped by a continuous laser with wavelength equal to $404$nm. One of the two photons, the signal, is sent along the apparatus, while the other is measured by a single photon detector and acts as a trigger for the experiment. The probe state is prepared by initializing the single-photon polarization in the linear horizontal state $\ket{H}$, through a polarizing beam splitter (PBS). After the PBS, the photon passes through a QP with a topological charge $q$ and a half-wave plate (HWP) which inverts its polarization, generating the following superposition:
\begin{equation}
    \ket{\Psi}_0=\frac{1}{\sqrt{2}}\big(\ket{R}\ket{+m}+\ket{L}\ket{-m}\big),
\end{equation}
where $m=2q$ is the value, in modulus, of the OAM carried by the photon. In this way, considering also the spin angular momentum carried by the polarization, the total angular momenta of the two components of the superposition are $\pm|m+1|$.

After the probe preparation, the generated state propagates and reaches the receiving station, where it enters in a measurement apparatus rotated by an angle $\theta$. Such a rotation is encoded in the photon state by means of a relative phase shift with a value $2 |m+1|\, \theta$ between the two components of the superposition:
\begin{equation}
    \ket{\Psi}_1=\frac{1}{\sqrt{2}}\big(e^{i\, (m+1) \theta}\ket{R}\ket{+m}+e^{-i\, (m+1) \theta}\ket{L}\ket{-m}\big) \;.
\end{equation}
To measure and retrieve efficiently the information on $\theta$, such a vector vortex state is then reconverted into a polarization state with zero OAM. This is achieved by means of a second HWP and a QP with the same topological charge as the first one, oriented as the rotated measurement station: 
\begin{equation}
    \ket{\Psi}_2=\frac{1}{\sqrt{2}}\big(\ket{R}+e^{-i\, 2 (m+1) \theta}\ket{L}\big) \;,
\label{eq:encodedState}
\end{equation}
where the zero OAM state factorizes and is thus omitted for ease of notation.
In this way, the relative rotation between the two apparatuses is embedded in the polarization of the photon in a state which, for $s=m+1$, exactly mimics~the output vector $|\Psi_s(\theta)\rangle$ of the previous section and that is finally measured with a PBS (concordant with the rotated station) followed by single photon detectors. Note that a HWP is inserted just after the preparation PBS and before the first three QPs. Such a HWP is rotated by $0^\circ$ and $22.5^\circ$ during the measurements to obtain the projections in the $|H\rangle$, $|V\rangle$ basis and in the diagonal one ($|D\rangle$, $|A\rangle$). In each stage, half of the photons are measured in the former basis, and half in the latter. The entire measurement station is mounted on a single motorized rotation cage. The interference fringes at the output of such a setup oscillates with an output transmission probability $P=\cos^2 [(m+1)\theta]$ with a periodicity that is $\pi/(m+1)$. Hence, the maximum periodicity is $\pi$ at $m=0$ and, consequently, one can unambiguously estimate at most all the rotations in the range $[0,\pi)$. 

The limit of the error on the estimation $\hat{\theta}$ of the rotation $\theta$ is:
\begin{equation}
    \Delta \hat{\theta}\ge \frac{1}{2\,(m+1)\sqrt{\nu\,n} }\;,
    \label{eq:subSQLbound}
\end{equation}
where $n$ is the number of the employed single photons carrying a total angular momentum $(m+1)$ and $\nu$ is the number of repetition of the measurement. Such a scaling is Heisenberg-like in the angular momentum resource $m+1$, and can be associated with the Heisenberg scaling achievable by multi-pass protocols for phase estimation, using non-entangled states \cite{higgins2007entanglement}. This kind of protocols can overcome the SQL scaling, that in our case reads $1/(2\, \sqrt{\nu\,n})$. However, such a limit can be achieved only in the asymptotic limit of $\nu \rightarrow \infty$,
where the scaling of the precision in the total number of resources used is again the classical one $\Delta \hat{\theta} \sim 1/\sqrt{N}$, if the angular momentum is not increased. Here, we investigate both the non-asymptotic and near asymptotic regime using non-adaptive protocols. Our apparatus is an all automatized toolbox generalizing the photonic gear presented in \cite{dambrosio_gear2013}. In our case, six QPs are simultaneously aligned in a cascaded configuration and actively participate in the estimation process. The first three QPs, each with a different topological charge $q$, lie in the preparation stage, while the other three, each having respectively the same $q$ of the first three, in the measurement stage. All the QPs are mounted inside the same robust and compact rotation stage able to rotate around the photon propagation direction. Notably, the whole apparatus is completely motorized and automated. Indeed, both the rotation stage and the voltages applied to the q-plates are driven by a computing unit which fully controls the measurement process.

During the estimation protocol of a rotation angle, only one pair of QPs with the same charge, one in the preparation and the other in the measurement stage, is simultaneously turned on. For a fixed value of the rotation angle, representing the parameter to measure, pairs of QPs with the same charge are turned on, while keeping the other pairs turned off. Data are then collected for each of the four possible configurations, namely all the q-plates turned off, i.e. $s=1$, and the three settings producing $s=2,11,51$, respectively. Finally, the measured events are divided among different estimation strategies and exploited for the post processing analysis. 

\section*{Results}
The optimization of the uncertainty on the estimated rotation angle is obtained by employing the protocol described above. In particular, such approach determines the use of the resources of each estimation stage. In this experiment, we have access to two different kinds of resources, namely the number of photon-pairs $n$ employed in the measurement and the value of their total angular momentum $s$. Therefore, the total number of employed resources is $N=\sum_{i=1}^K n_is_i$, where $n_i$ is the number of photons with momentum $s_i$, and $K = 4$. According to the above procedure, for every $N$ we determine the sequence of the multiplicative factors $s_i$ and $n_i$ associated to the optimal resource distribution.

The distance between the true value, $\theta$, and the one obtained with the estimation protocol, $\hat\theta$, is obtained computing the circular error as follows:
\begin{equation}
    |\hat\theta-\theta| = \frac{\pi}{2}- \Big |(\theta-\hat\theta) \text{ mod }\pi - \frac{\pi}{2} \Big| \;.
\end{equation}
Repeating the procedure for $r = 1,\dots, R$ different runs of the protocol with $R=200$, we retrieve, for each estimation strategy, the corresponding root mean square error (RMSE):
\begin{equation}
    \Delta\hat\theta = \sqrt{\sum_{r=1}^{R}\frac{|\hat\theta_i-\theta|^2}{R}}.
\end{equation}
 We remark that $R$ and $\nu$ in Eq.~\eqref{eq:subSQLbound} do not have  the same interpretation. Indeed $R$ is not a part of the protocol, but is merely the number of times we repeat it in order to get a reliable estimate of its precision. We then averaged such quantity over $17$ different rotations with values between $0$ and $\pi$, leading to $\overline{\Delta\hat\theta}$. In such a way, we investigate the uncertainty independently on the particular rotation angle inspected.
\begin{figure}[h]
  \includegraphics[width=\columnwidth]{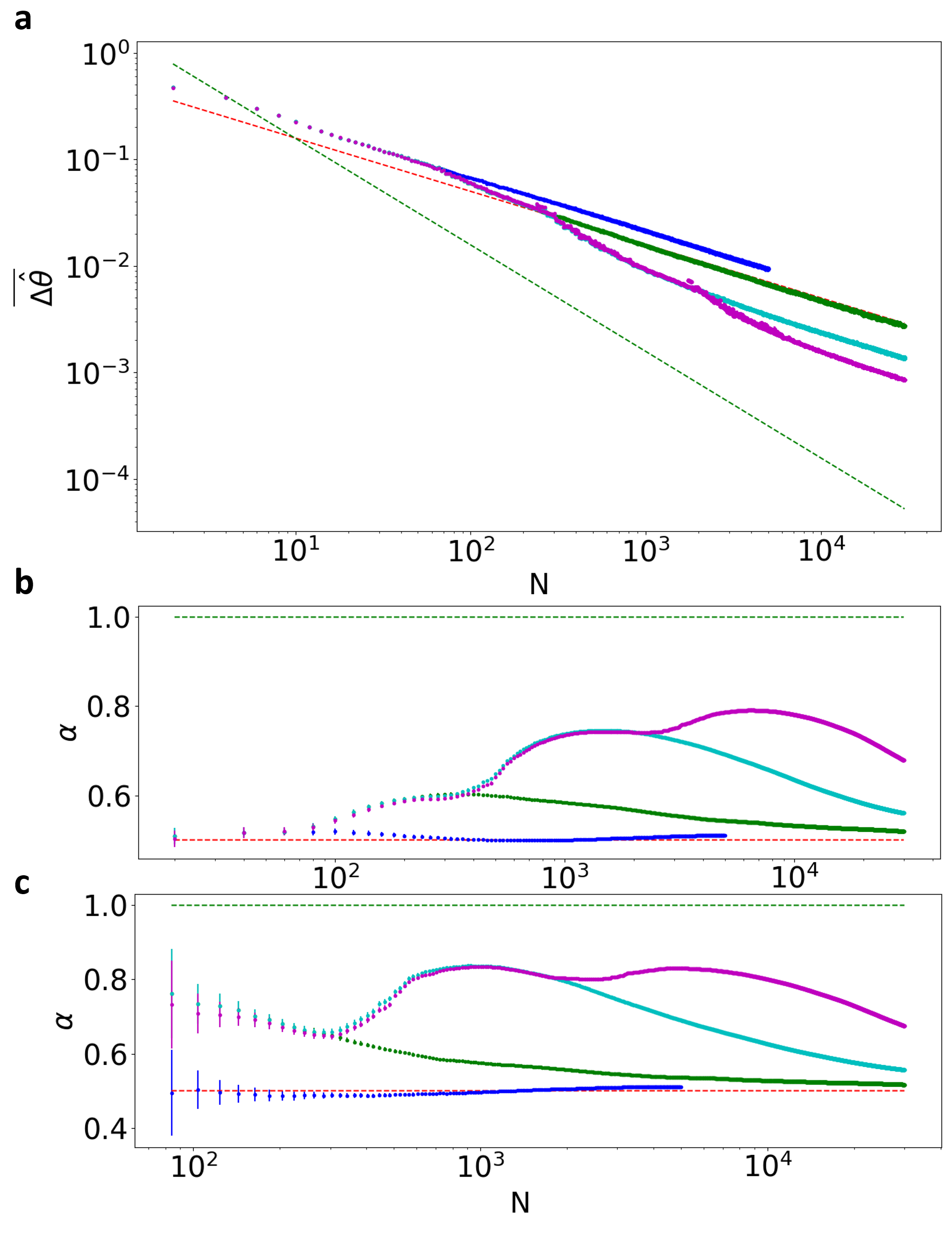}
    \caption{{\bf Approaching the HL with higher-order OAM states.} \textbf{a)} Averaged measurement uncertainty over $R=200$ repetitions of the algorithm and over $17$ different angle measurements, in the interval $[0,\pi)$, as a function of the total amount of resources $N$. The adoption of single-photon states with progressively higher-order total angular momentum allows to progressively approach the HL. The red dashed line is the standard quantum limit for this system $1/(2\sqrt{N})$, while the green dashed line is the HL $\pi/(2N)$. \textbf{b)} Value of the coefficient $\alpha$ and its standard deviation obtained by fitting the points from $N=2$ to the value reported on the $x-$axis with the curve $C/N^\alpha$. \textbf{c)} Value of the coefficient $\alpha$ and its standard deviation obtained by fitting the points from $N=N_0$ to the value reported on the $x-$axis with the curve $C/N^\alpha$. Purple points: estimation process with the full strategy. Blue points: estimation process by using only $s = 1$. Green points: estimation process by using only $s = 1; 2$. Cyan points: estimation by using only $s = 1; 2; 11$.}
    \label{fig:rmse}
\end{figure}
\begin{figure*}[t]
  \includegraphics[width=\textwidth]{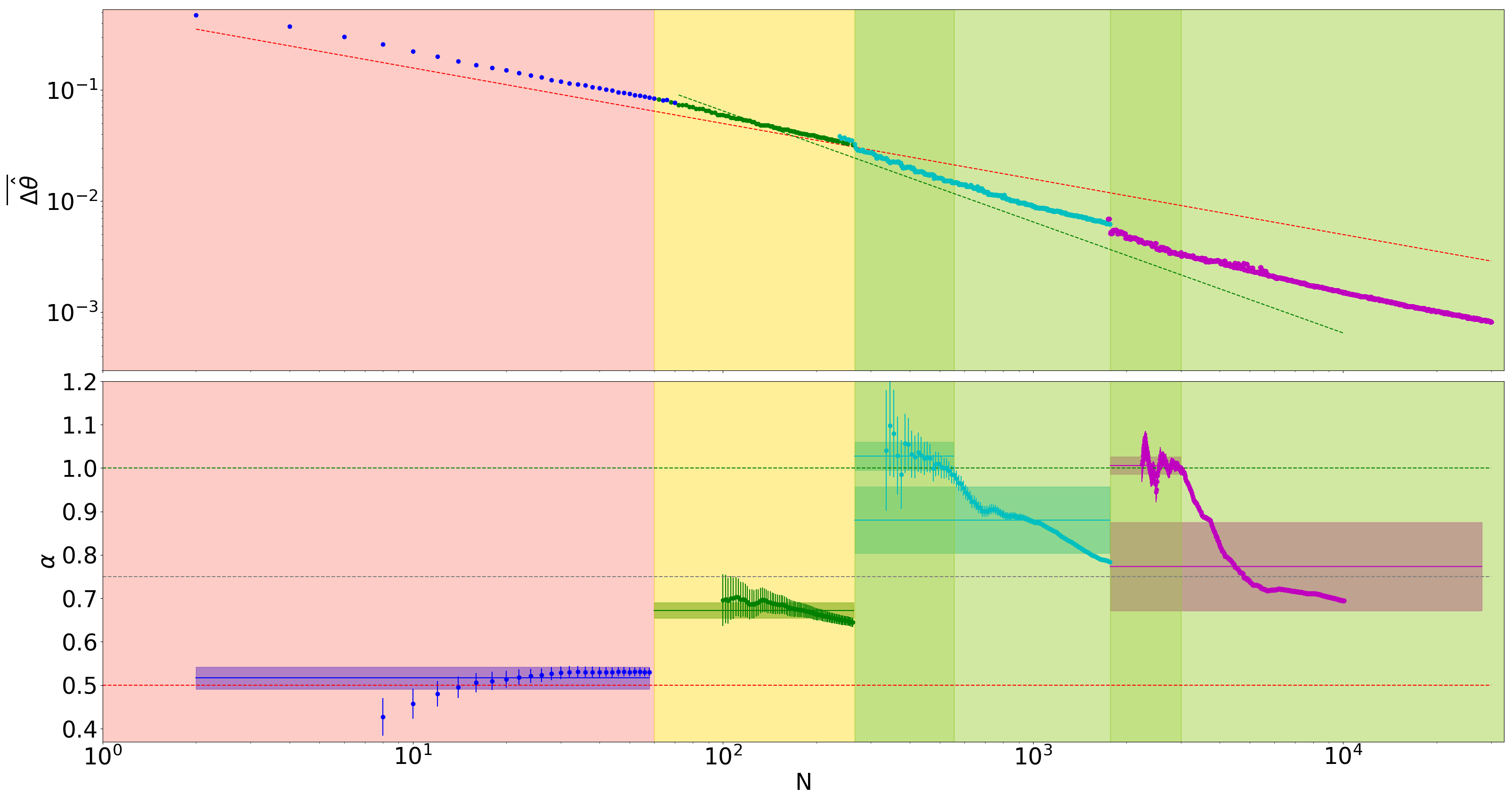}
    \caption{{\bf Certification of the Heisenberg scaling in the local scenario.} Upper panel: measurement uncertainty averaged over $17$ different angle values in the interval $[0,\pi)$ as a function of the amount of resources $N$. We highlight the points with the color code associated to the maximum value of $s$ exploited in each strategy. Blue points: strategies with $s = 1$. Green points: strategies relative to $s = 1; 2$. Cyan points: strategies relative to $s = 1; 2; 11$. Purple points: strategies for $s = 1; 2; 11; 51$. Error bars are smaller than the size of each point. 
    Lower panel: value of the coefficient $\alpha$ and the relative confidence interval for the four inspected regions. Such a confidence interval consists in a $3 \sigma$ region, obtained for the best fit with function $C/N^\alpha$. The fit is done on batches of data as described in the main text. The continuous lines show the average value of $\alpha$ in the respective region, while the shaded area is its standard deviation.
    In both the plots the salmon, yellow and green colored areas represent respectively region with  SQL scaling ($\alpha=0.5$), sub-SQL scaling ($0.5<\alpha\le 0.75$) and a scaling approaching the Heisenberg-limit ($0.75<\alpha\le 1$). The red dotted line represents the SQL $= 1/(2\sqrt{N})$ ($\alpha = 0.5$) while the green one is the HL $=C/(2N)$ ($\alpha = 1$). The grey dotted line is the threshold $\alpha = 0.75$.}
    \label{fig:rmse2}
\end{figure*}
 In the following we report the results of our investigation on how the measurement sensitivity is improved by exploiting strategies that have access to states with an increasing value of the total angular momentum, obtained by tuning QPs with higher topological charge. We first consider the scenario where only photon states with $s=1$ are generated. In this case, the RMSE follows as expected the SQL scaling as a function of the number of total resources. The obtained estimation error for the strategies constrained by such condition is represented by the blue points in Fig.~\ref{fig:rmse}a. Running the estimation protocol and exploiting also states with $s>1$ it is possible to surpass the SQL and progressively approach Heisenberg-limited performances, for high values of $s$. In particular, we demonstrate such improvement by progressively adding to the estimation process a new step with higher OAM value. We run the protocol limiting first the estimation strategy to states with $s = 1; 2$ (green points), then to $s = 1; 2; 11$ (cyan points) and finally to $s = 1; 2; 11; 51$ (magenta points). For each scenario, the number of photons $n$ per step is optimized accordingly. Performing the estimation with all the $4$ available orders of OAM allows us to achieve an error reduction, in terms of the obtained variance, up to $10.7$ dB below the SQL. Note that the achievement of the Heisenberg scaling is obtained by progressively increasing the order of the OAM states employed in the probing process, mimicking the increase of $N$ when using N00N-like states in multipass protocols. This is highlighted by a further analysis performed in Fig.~\ref{fig:rmse}b and Fig.~\ref{fig:rmse}c. More specifically, if beyond a certain value of $N$ the OAM value is kept fixed, the estimation process will soon return to scale as the SQL. 

To certify the quantum-inspired enhancement of the sensitivity scaling, we performed a first global analysis on the uncertainty scaling considering the full range of $N$. This is performed by fitting the obtained experimental results with the function $C/N^\alpha$. In particular, such a fitting procedure is performed considering batches of increasing size of the overall data. This choice permits to investigate how the overall scaling of the measurement uncertainty, quantified by the coefficient $\alpha$, changes as function of $N$. Starting from the point $N=2$ we performed the fit considering each time the subsequent $10$ experimental averaged angle estimations (reported in Fig.\ref{fig:rmse}a), and evaluated the scaling coefficient $\alpha$ with its corresponding confidence interval for each data batch. The results of this analysis are reported in Fig.~\ref{fig:rmse}b. As shown in the plot, $\alpha$ is compatible with the SQL, i.e. $\alpha=0.5$, when the protocol employs only states with $s=1$. Sub-SQL performance are conversely achieved when states with $s>1$ are introduced in the estimation protocol. The scaling coefficient of the best fit on the experimental data collected when exploiting all the available QPs (magenta points) achieves a maximum value of $\alpha=0.7910\pm0.0002$, corresponding to the use of $6,460$ resources. The enhancement is still verified when the fit is performed considering the full set of $30,000$ resources. Indeed, the scaling coefficient value in this scenario still remains well above the SQL, reaching a value of $\alpha=0.6786\pm0.0001$. Given that the data sets corresponding to $s = 1$ inherently follow the SQL, we now focus on those protocols with $s>1$, thus taking into account only points starting from $N_0 = 62$. This value coincides with the first strategy exploiting states with $s=2$. Fitting only such region the maximum value of the obtained coefficient increases to $\alpha=0.8301\pm0.0003$ for $N=4,764$. Note that, as higher resource values $s$ are introduce, the overall scaling coefficient of the estimation process, taking into account the full data set, progressively approaches the value for the HL.

Then, we focus on the protocols which have access to the full set of states with $s = 1; 2; 11; 51$, and we perform a local analysis of the scaling, studying individually the regions defined by the order of OAM used, and characterized by different colors of the data points in the top panel of Fig.~\ref{fig:rmse2}. This is performed by fitting the scaling coefficient with a batch procedure (as described previously) within each region. We first report in the top panel of Fig.~\ref{fig:rmse2} the obtained uncertainty $\overline{\Delta \hat{\theta}}$. Then, we study the overall uncertainty scaling which shows a different trend depending on the maximum $s$ value we have access to. To certify locally the achieved scaling, we study the obtained coefficient for the four different regions sharing strategies requiring states with the same maximum value of $s$. In the first region ($2 \le N \le 60$), since $s = 1$ no advantage can be obtained respect to the SQL. This can be quantitatively demonstrated studying the compatibility, in $3\sigma$, of the best fit coefficient $\alpha$ with $0.5$. Each of the blue points in the lower panel of Fig.~\ref{fig:rmse2} is indeed compatible with the red dashed line. In the second region ($62 \le N \le 264$), since states with $s=2$ are also introduced, it is possible to achieve a sub-SQL scaling. When states with up to $s=11$ and $s=51$ are also employed ($N > 264$) we observe that the scaling coefficient $\alpha>0.75$ is well above the value obtained for the SQL.  Finally, we can identify two regions ($266 \le N \le 554$ and $1,772 \le N \le 2,996$) where the scaling coefficient $\alpha$ obtained from a local fit is compatible, within $3\sigma$, with the value $\alpha = 1$ corresponding to the exact HL. This holds for extended resource regions of size $\sim 300$ and $\sim 1,000$, respectively, and provides a quantitative certification of the achievement of Heisenberg-scaling performances.

\section*{Discussion and Conclusion}
The achievement of Heisenberg precision for a large range of resources $N$ is one of the most investigated problems in quantum metrology. Recent progress have been made demonstrating a sub-SQL measurement precision approaching the Heisenberg limit when employing a restricted number of physical resources. However, beyond the fundamental purpose of demonstrating the effective realization of a Heisenberg limited estimation precision, it becomes crucial for practical applications to maintain such enhanced scaling for a sufficiently large range of resources.

We have experimentally implemented a protocol which allows to estimate a physical parameter with a Heisenberg scaling precision in the non-asymptotic regime. In order to accomplish such a task, we employ single-photon states carrying high total angular momentum generated and measured in a fully automatized toolbox using a non-adaptive estimation protocol. Overall, we have demonstrated a sub-SQL scaling for a large resource interval $O(30,000)$, and we have validated our results with a detailed global analysis of the achieved scaling as function of the employed resources. Furthermore, thanks to the extension of the investigated resource region  and to the abundant number of data points, we can also perform a local analysis which quantitatively proves the Heisenberg scaling in a considerable range of resources $O(1,300)$. This represents a substantial improvement over the state of the art of the Heisenberg scaling protocols.

These results provide experimental demonstration of a solid and versatile protocol to optimize the use of resources for the achievement of quantum advantage in \emph{ab-initio} parameter estimation protocols. Given that its use can be adapted to different platforms and physical scenarios, this opens new perspectives to achieve Heisenberg scaling for a broad resource value. Direct near-term applications of the methods can be foreseen in different fields including sensing, quantum communication and information processing. 



\section*{Acknowledgments}
This work is supported by the ERC Advanced grant PHOSPhOR (Photonics of Spin-Orbit Optical Phenomena; Grant Agreement No. 828978), by the Amaldi Research Center funded by the  Ministero dell'Istruzione dell'Universit\`a e della Ricerca (Ministry of Education, University and Research) program ``Dipartimento di Eccellenza'' (CUP:B81I18001170001) and by MIUR (Ministero dell’Istruzione, dell’Università e della Ricerca) via project PRIN 2017 “Taming complexity via QUantum Strategies a Hybrid Integrated Photonic approach” (QUSHIP) Id. 2017SRNBRK.


%

\section*{Methods}
\subsection*{Experimental details}
Generation and measurement of the OAM states is obtained through the q-plate devices. A q-plate is a tunable liquid crystal birefringent element that couples the polarization and the orbital angular momentum of the incoming light. Given an incoming photon carrying an OAM value $l$, the action of a tuned QP in the circular polarization basis ${\ket{R}, \ket{L}}$ is the following transformations:
\begin{equation}
\label{eq:qplatepi}
\begin{split}
QP\, \ket{R}_{\pi} \ket{l}_{oam}=\ket{L}_{\pi} \ket{l-2q}_{oam},\\
QP\, \ket{L}_{\pi} \ket{l}_{oam}=\ket{R}_{\pi} \ket{l+2q}_{oam},
\end{split}
\end{equation}
where $ \ket{X}_{\pi}$ indicates the polarization $X$, while $\ket{l\pm 2q}_{oam}$ indicates the OAM value $\pm 2q$, being $q$ is the topological charge characterizing the QP. Tuning of the q-plate is performed by changing the applied voltage.

The two sources of error are photon loss and non-unitary conversion efficiency of the QPs. Each stage of the protocol is characterized by a certain photon loss $\eta$, which reduces the amount of detected signal. In general each stage will have its own noise level $\eta_j$. Conversely, non-unitary efficiency of the QPs translates to a non-unitary visibility $v$, which changes the probability outcomes for the two measurement projections as following:
\begin{equation}
\begin{split}
    p_{HV} &= \frac{1}{2}\big(1+v\cos{2s\theta}\big),\\
    p_{DA} &= \frac{1}{2}\big(1+v\sin{2s\theta}\big).
    \label{eq:prob}
 \end{split}   
\end{equation} 

The limit of the error on the estimation of the rotation $\theta$ in this setting is:
\begin{equation}
    \Delta \hat{\theta}\ge \frac{1}{2\,(m+1)v\sqrt{\nu\eta\,n} }\;
\end{equation}
However, if the actual experiment is performed in post-selection, then $\eta=1$ and the only source of noise is the reduced visibility.

\subsection*{Angle error dependence}
\begin{figure}[h]
  \includegraphics[width=\columnwidth]{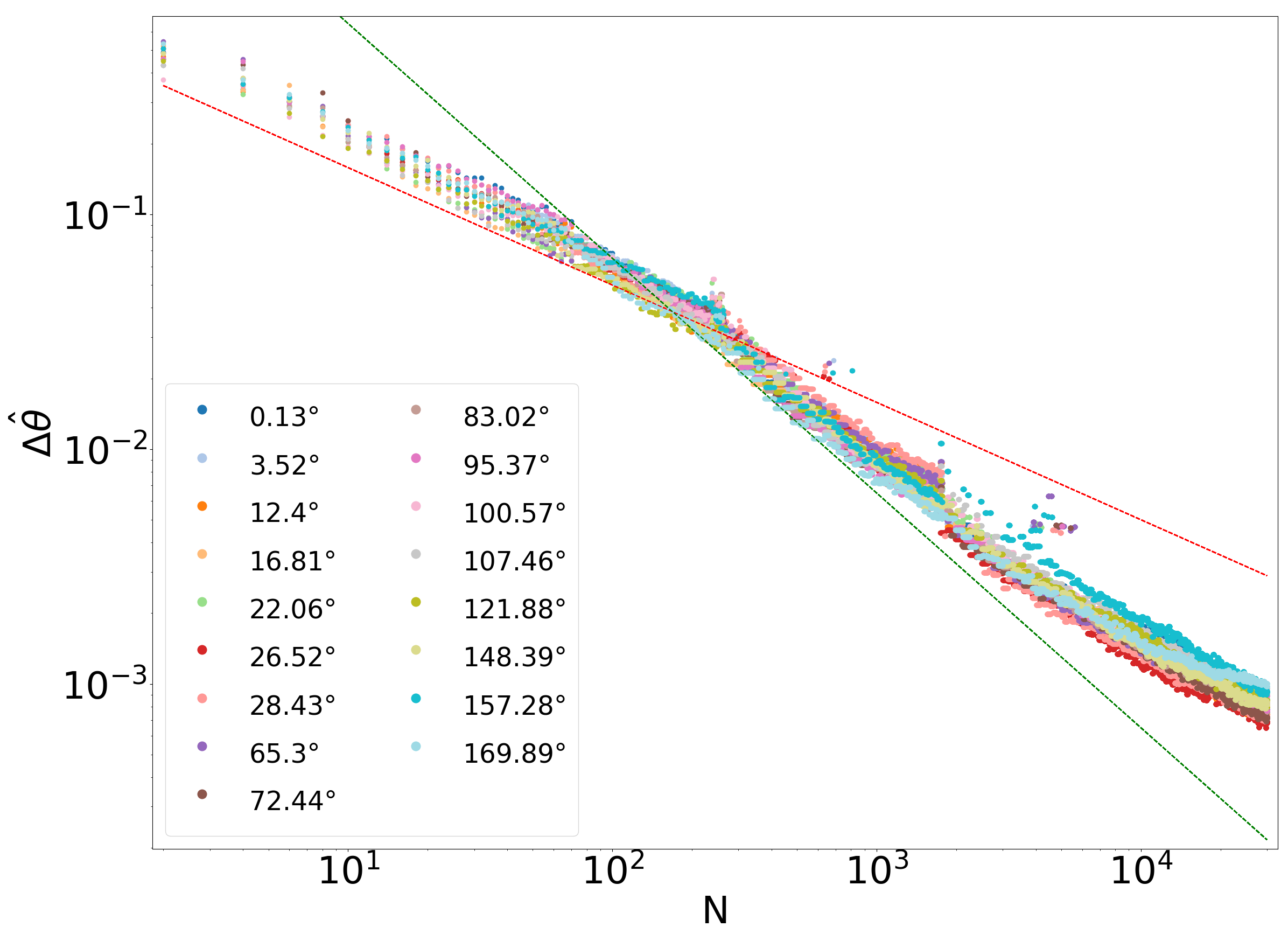}
    \caption{RMSE obtained in $200$ runs of the estimation protocol for each of the $17$ angles measured.}
    \label{fig:angoli}
\end{figure}

\begin{figure}[h]
  \includegraphics[width=\columnwidth]{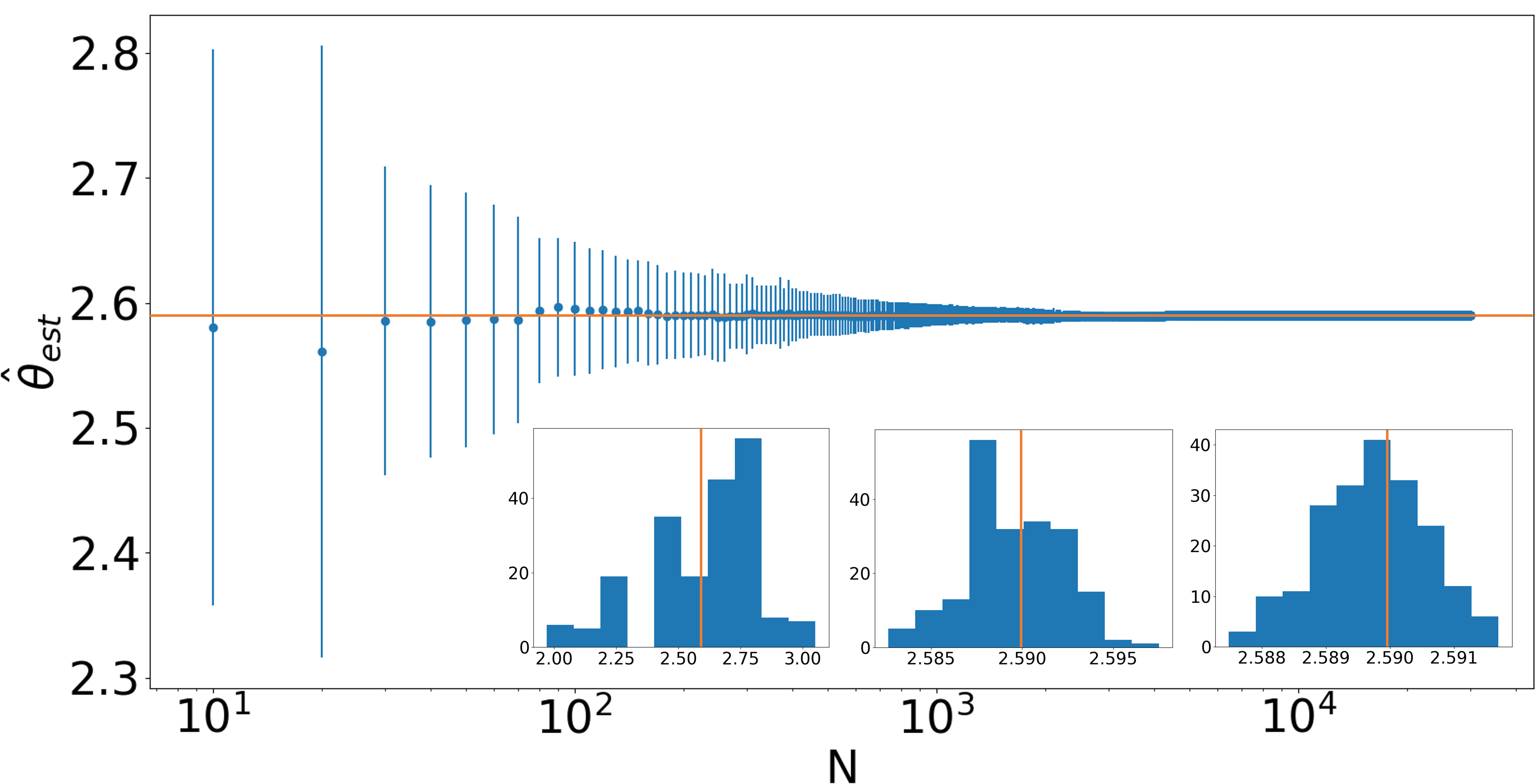}
    \caption{Value of the estimated angle $\hat{\theta}$ at each iteration of the protocol. The orange line shows the true value $\theta = 2.59$ rad. Inset: histograms of the obtained value in the 200 different repetition of the protocol for $N=10; N=5,000$ and $N=30,000$ respectively.}
    \label{fig:convergence}
\end{figure}

\begin{figure}[h]
  \includegraphics[width=0.9\columnwidth]{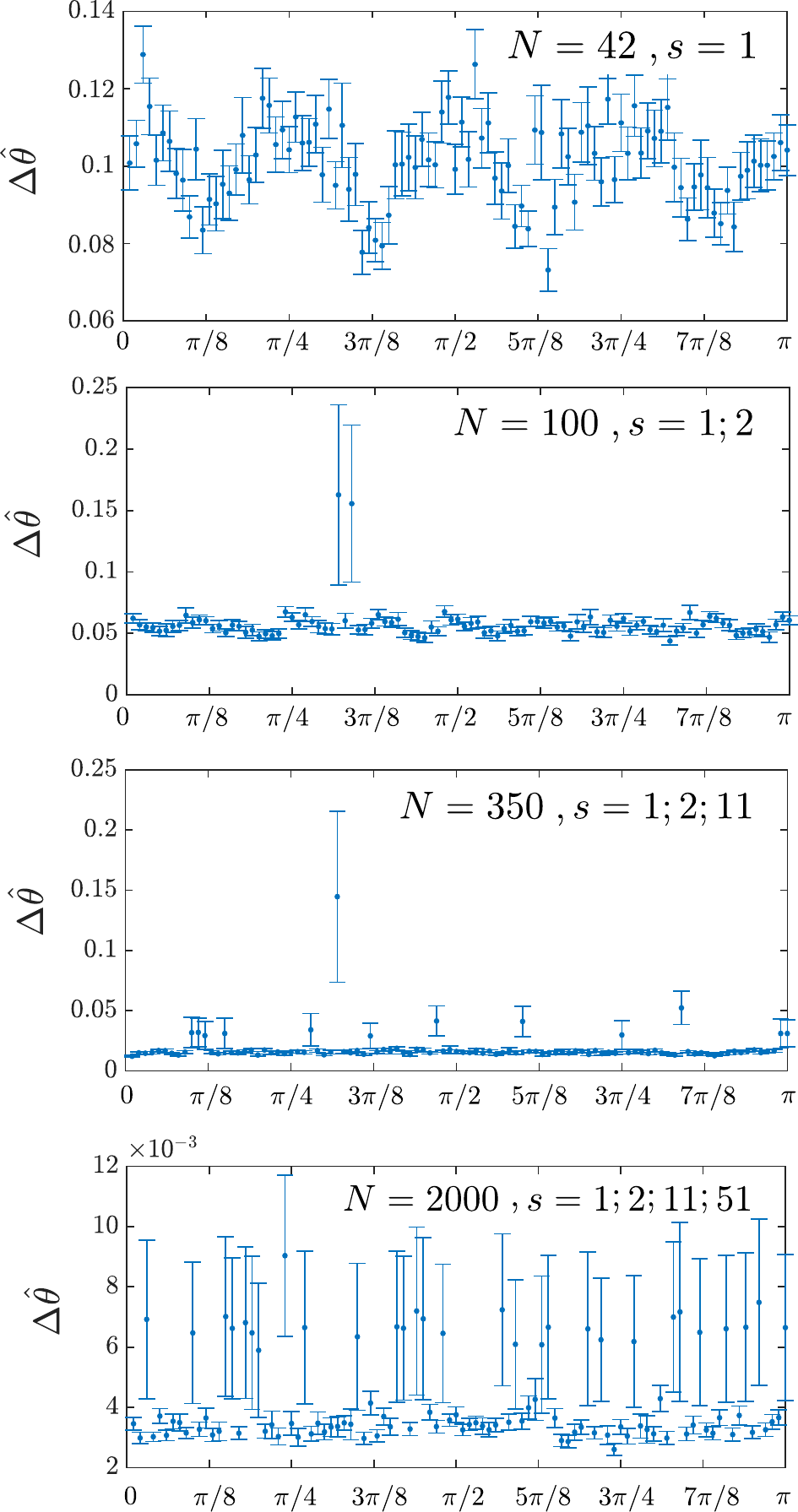}
    \caption{Simulation of the error in the four stages of the estimation for representative values of $N$ in the sub-SQL regime as a function of the angle $\theta$. The algorithm is non-adaptive, and a periodical dependence of the error as a function of the angle can be recognised in the first plot, where no quantum resources are used. In the subsequent stages the error peaks correspond to the experiments in which a wrong interval has been chosen. These outliers are a product of the statistical nature of the measurements and their position is not associated to a particular $\theta$. Despite the error being a fluctuating quantity, an arbitrary uniform sample of angles is sufficient to characterize its average behaviour. For each point in the plot $100$ experiment has been simulated.}
    \label{fig:thetaStages}
\end{figure}

To verify the independence of the achieved precision from the particular rotation value measured experimentally we applied the protocol to estimate $17$ different rotation angles, nearly uniformly distributed in the interval $[0,\pi)$. The robustness of the implemented protocol over the choice of the particular values inspected is demonstrated experimentally looking at the results obtained for each of the estimated angle reported in Fig.~\ref{fig:angoli}. Further insight on this aspect is obtained by verifying the convergence of the estimation protocol to the true value $\theta$, shown by way of example in Fig.~\ref{fig:convergence} for one of the $17$ inspected angles.
Such independency is also confirmed by studying the protocol performances on simulated data as a function of the value of the rotation angle $\theta$ inspected. From Fig.~\ref{fig:thetaStages} we can deduce that in the sub-SQL regions the error is dominated by random fluctuations, where the outliers correspond to errors in the localization procedure and do not show correlation with $\theta$.

\subsection*{Details on fitting results}
The best fits obtained considering the data points for each of the different regions of Fig.~\ref{fig:rmse2} are reported in Table \ref{tab}.
\begin{table}[htp]
\centering
\begin{tabular}{|c|c|c|}
\hline
N & $\alpha$ & $R^2$ \\
\hline
\hline
$2 - 60$ & $0.530 \pm 0.003$ & $99.0\%$ \\
$62 - 264$ & $0.645 \pm 0.003$ & $99.2\%$ \\
\boldsymbol{$266 - 554$} & \boldsymbol{$0.984 \pm 0.007$} & \boldsymbol{$98.7\% $}\\
$556 - 1,770$ & $0.727 \pm 0.002$ & $98.7\%$ \\
\boldsymbol{$1,772 - 2,996$} & \boldsymbol{$0.995 \pm 0.004$} &  \boldsymbol{$96.4\%$} \\
$2,998 - 30,000$ & $0.601 \pm 0.001$ & $99.3\%$ \\
\hline
\end{tabular}
\caption{Values of the best fit and relative $R^2$ for the experimental data using the power law $\propto 1/N^\alpha$, performed for different intervals of employed resources. In bold are reported the regions with Heisenberg scaling, certified by a value of $\alpha$ compatible with $1$.}
\label{tab}
\end{table}
A more detailed study of the error dependence on the value of $\theta$ showed that some outliers can emerge in the overall low RMSE achieved with our protocol. The outliers arise from a wrong assessment of the rotation interval selected in the early stages of the protocol. Averaging the obtained RMSE for each single angle over different repetitions of the protocol allows to considerably reduce the presence of such spikes (see Fig.~\ref{fig:angoli}). On the contrary, the outliers are almost completely mitigated averaging the performances also over the different angle values measured. This can be indeed verified looking at the top panel of Fig.~\ref{fig:rmse2} in the main text. To study locally the error scaling indeed we performed the fit starting from a small batch of experimental measured point. Although the outliers are negligible in the overall trend their influence can mislead from a reliable analysis of the error trend when considering the small batches over which we perform the local analysis. Therefore, in order to completely wash out their influence when studying the achieved scaling we removed all the outliers present in the curve by looking at the residual value.

Note that the fitting procedure on the experimental points is weighted with their corresponding error bar. The error bars on the RMSE values averaged over the different angles inspected and reported in Fig.~\ref{fig:rmse} have been obtained with the following procedure. For each angle $j =1,\dots,J$ with $J=17$, the error associated to the different strategies is obtained considering the standard deviation over the multiple repetitions $r=1,\dots,R$ of the estimation protocol, as follows:
\begin{equation}
    \delta(\overline{\Delta\hat\theta}) = \frac{1}{J} \sqrt{\sum_{j=1}^{J} \text{Var}\big(\Delta\hat\theta_j\big)}.
\end{equation}
Fixing the angle $j$ the variance of the RMSE over the $R=200$ repetitions is obtained propagating the averaged error of the absolute distance between the estimated angle and the true one:  
\begin{equation}
   \text{Var}\big(\Delta\hat\theta_j\big) = \frac{1}{2} \frac{\text{Var}\big(|\hat{\theta}_{j}-\theta_j|^2\big)}{\sqrt{\sum_{r=1}^{R}|\hat{\theta}_{rj}-\theta_j|^2}},
\end{equation}
where $\hat{\theta}_{rj}$ is the $r$-th estimated for angle $j$, and $\theta_j$ is the true value of angle $j$. Combining the two formulas we obtain the expression used to compute the error bars reported in the plots and used to perform the weighted fit:
\begin{equation}
    \delta(\overline{\Delta\hat\theta}) = \frac{1}{J} \sqrt{\sum_{j=1}^{J} \frac{1}{2} \frac{\text{Var}\big(|\hat{\theta}_{j}-\theta_j|^2\big)}{\sqrt{\sum_{r=1}^{R}|\hat{\theta}_{rj}-\theta_j|^2}}}.
\end{equation}

\subsection*{Data processing algorithm and optimization}
In this section we present extensively the phase estimation algorithm which we used to process the measured data, and its optimization. As it naturally applies to a phase in $\left[0, 2 \pi \right)$ we present it for $\varphi = 2 \theta \in \left[0, 2 \pi \right)$. At each stage of the procedure the estimator $\hat{\varphi}$ and its error $\Delta \hat{\varphi}$ can be easily converted in estimator and error for the rotation angle: $\hat{\theta} = \hat{\varphi}/2$ and $\Delta \hat{\theta} = \Delta \hat{\varphi}/2$. In the $i$-th stage of the procedure we are given the result of $n_i/2$ photon polarization measurements on the basis $HV$ and $n_i/2$ measurements on the basis $DA$. We define $\hat{f}_{HV}$ and $\hat{f}_{DA}$ the observed frequencies of the outcomes $H$ and $D$ respectively and introduce the estimator $\widehat{s_i \theta} = \text{atan2} ( 2 \hat{f}_{HV} -1, 2 \hat{f}_{DA} -1 ) \in [ 0, 2 \pi )$. From the probabilities in~\eqref{eq:prob} it is easy to conclude that $\widehat{s_i \varphi}$ is a consistent estimator of $s_i \varphi \mod 2 \pi$. This does not identify an unambiguous $\varphi$ alone though, but instead a set of $s_i$ possible values $\widehat{s_i \varphi}/s_i + 2 \pi m/s_i$ with $m = 0, 1, 2, \cdots, s_i-1$. Centered around this points we build intervals of size $2 \pi/(s_i \gamma_i)$, where
\begin{equation}
	\gamma_i = \frac{\gamma_{i-1}}{\gamma_{i-1} - \frac{s_i}{s_{i-1}}} \; .
	\label{eq:generator}
\end{equation}
The algorithm then chooses among this intervals the only one that overlaps with the previously selected interval. The choice of $\gamma_i$, computed recursively with the formula in Eq.~\eqref{eq:generator}, is fundamental in order to have one and only one overlap. The starting point $\gamma_1$ of the recursive formula can be chosen freely inside an interval of values that guarantees $\gamma_i \ge 1 \; \forall i$, therefore it will be subject to optimization. By convention we set $\gamma_0 = 1$. The Algorithm~\ref{alg:algorithm1} reports in pseudocode the processing of the measurement outcomes required to get the estimator $\hat{\varphi}$ working at Heisenberg scaling.
\begin{algorithm}[H]
	\caption{Phase estimation}
	\label{alg:algorithm1}
	\begin{algorithmic}[1]
		\State $\hat{\varphi} \gets 0$
		\For {$i = 1 \to K$}
		\State $\left[ 0, 2 \pi \right) \ni \widehat{s_i \varphi} \gets$ Estimated from measurements.
		\State $\left[ 0, \frac{ 2 \pi}{s_i} \right) \ni \hat{\xi} \gets \frac{\widehat{s_i \varphi}}{s_i}$
		\State $m \gets \Big \lfloor \frac{s_i \hat{\varphi}}{2 \pi} - \frac{1}{2} \frac{s_i}{s_{i-1} \gamma_{i-1}} \Big \rfloor$
		\State $\hat{\xi} \gets \hat{\xi} + \frac{2 \pi m}{s_i}$
		\If {$\hat{\varphi} + \frac{\pi ( 2 \gamma_i -1 )}{s_i \gamma_i} - \frac{\pi}{s_{i-1} \gamma_{i-1}} < \hat{\xi} < \hat{\varphi} + \frac{\pi ( 2 \gamma_i+1 )}{s_i \gamma_i} + \frac{\pi}{s_{i-1} \gamma_{i-1}}$}
		\State $\hat{\varphi} \gets \hat{\xi} - \frac{2 \pi}{s_i}$
		\ElsIf {$\hat{\varphi} - \frac{\pi ( 2 \gamma_i+1 )}{s_i \gamma_i} - \frac{\pi}{s_{i-1} \gamma_{i-1}} < \hat{\xi} < \hat{\varphi} - \frac{\pi ( 2 \gamma_i-1 )}{s_i \gamma_i} + \frac{\pi}{s_{i-1} \gamma_{i-1}}$}
		\State $\hat{\varphi} \gets \hat{\xi} + \frac{2 \pi}{s_i}$
		\Else
		\State $\hat{\varphi} \gets \hat{\xi}$
		\EndIf
		\State $\hat{\varphi} \gets \hat{\varphi} - 2 \pi \lfloor \frac{\hat{\varphi}}{2 \pi} \rfloor$
		\EndFor
	\end{algorithmic}
\end{algorithm}
We can upper bound the probability of choosing the wrong interval through the probability for the distance of the estimator $\widehat{s_i \varphi}$ from $\varphi$ to exceed $\pi/ \gamma_i$, that is
\begin{equation}
    \text{P} [ | \widehat{s_i \varphi} - \varphi| \ge \frac{\pi}{\gamma_i} ] \le A C(\gamma_i)^{-\frac{n_i}{2}} \; ,
    \label{eq:probUpp}
\end{equation}
where $n_i$ is the number of photons employed in the stage, $C(\gamma) = \exp \left[ b \sin^2 \left( \frac{\pi}{\gamma} \right) \right]$, and $A$ is an unimportant numerical constant. This form for $C (\gamma)$ was suggested by the Hoeffding's inequality, and we set $b = 0.7357$ as indicated by numerical evaluations for $n_i \le 40$. By applying the upper bound in Eq.~\eqref{eq:probUpp} we could write a bound on the precision of the final estimator $\hat{\varphi}$, as measured by the RMSE with the circular distance, that reads
\begin{multline}
	\Delta^2 \hat{\varphi} \le \frac{A \pi^2}{2 b n_K s_K^2} + \frac{3 A \pi^2}{4 s_K^2} e^{-\frac{b n_K}{2}}  + \\ + \sum_{i=1}^{K-1} \left( \frac{2 \pi D_i}{\gamma_{i-1} s_{i-1}}\right)^2 A C_i^{-\frac{n_i}{2}} \; .
	\label{eq:upperBoundGeneral}
\end{multline}
where $C_i = C \left( \gamma_i \right)$, and $D_i$ are
\begin{equation}
	D_i := \begin{cases} 
	\frac{1}{2} \; , & i = 1 \; ,\\
	1 + \gamma_{i-1} s_{i-1} \left[ \left( \sum_{k=i}^{K-2} \frac{1}{\gamma_k s_k} \right) + \right. \\ \left. \quad + \frac{1}{2 s_{K-1} \gamma_{K-1}} + \frac{1}{2 s_K} \right] \; , & i > 1 \; .
	\end{cases}
	\label{eq:expressionD}
\end{equation}
If $\frac{2 \pi D_i}{\gamma_{i-1} s_{i-1}} \ge \pi$ then we redefine $D_i = \frac{\gamma_{i-1} s_{i-1}}{2}$. The last stage of the estimation is different from the previous ones, as it is no more a step of the localization process. This difference can be clearly seen in how the error contribution is treated in Eq.~\eqref{eq:upperBoundGeneral}. We optimize this upper bound while fixing the total number of used resources by writing
\begin{multline}
	\mathcal{L} := \frac{\pi^2}{2 b n_K s_K^2} + \frac{3 \pi^2}{4 s_K^2} e^{-\frac{b n_K}{2}} + \sum_{i=1}^{K-1} \left( \frac{2 \pi D_i}{\gamma_{i-1} s_{i-1}}\right)^2 C_i^{-\frac{n_i}{2}} \\ - \lambda \left( \sum_{i=1}^{K} s_i n_i - N \right)\; .
	\label{eq:lagrangian}
\end{multline}
Through the optimization of this Lagrangian we found the resource distribution $n_i$ optimal for the given sequence of $s_i$ and $N$. Substituting back the obtained $n_i$ in the error expression we get
\begin{equation}
	\Delta^2 \hat{\varphi} \le \frac{A \pi^2}{2 b n_K s_K^2} + \frac{3 A \pi^2}{4 s_K^2} e^{-\frac{b n_K}{2}} + A e^{\alpha} \sum_{i=1}^{K-1} \frac{s_i}{\gamma_{i-1}^2 \log C_i} \; ,
	\label{eq:errorAlpha}
\end{equation}
where $\alpha$ depends on the total resource number $N$. In an experiment we have at disposal, or we have selected, a certain sequence of quantum resources $s = 1; s_2; s_3; \dots; s_K$, but it is not convenient for every $N$ to use the whole sequence. A better strategy is to add one at a time a new quantum resource as the total number of available resources $N$ grows, and therefore slowly building the complete sequence. For small $N$ we do not employ any quantum resource, so that $s=1$. The first upgrade prescribes the use of the $2$-stage strategy $s=1; s_2$, then, as $N$ reaches a certain value we upgrade again to a $3$-stage strategy $s=1;s_2;s_3$, and so on until we get to $s = 1; s_2; s_3; \dots; s_K$, which will be valid asymptotically in $N$. The optimal points at which these upgrades should be performed can be found by comparing the error upper bounds given in Eq.~\eqref{eq:errorAlpha} or via numerical simulations. The sequence $s=1; s_2; s_3; \dots, s_K$ might not be the complete set of all the quantum resources that are experimentally available. In our experiment $s = 1; 2; 11; 51$ were all the available quantum resources, but the here described procedure of adding one stage at a time might work better with only a subset of the available $s_i$. We are therefore in need of comparing many sets of quantum resources $s= 1; s_2; s_3; \dots; s_K$. The numerical simulations suggested us that a comparison of the summations
\begin{equation}
    \sum_{i=1}^{K-1} \frac{s_i}{\gamma_{j-1}^2 \log C_i} \; ,
\end{equation}
with optimized $\gamma_1$, that appear in Eq.~\eqref{eq:errorAlpha}, is a quick and reliable way to establish the best set of quantum resources. We can treat the non perfect visibility of the apparatus by rescaling the parameters $b$ and $C_j$ in the Lagrangian~\eqref{eq:lagrangian}. Given $v_i$ the visibility in the $i$-th stage, the rescaling requires $C_i \rightarrow C_i^{v_i^2}$, and for the last stage $b \rightarrow b v_K^2$.

\end{document}